# Simple test for high $J_c$ and low $R_s$ superconducting thin films

A. Sundaresan[a], Y. Tanaka[a], A. Iyo[a], M. Kusunoki[b] and S. Ohshima[b]

[a]Nanoelectronics Research Institute of AIST, AIST Tsukuba Central 2, 1-1-1 Umezono, Tsukuba, Ibaraki, 305 8568 Japan and CREST, JST, Saitama, 332-0012 Japan.

[b]Department of Electrical and information Engineering, Yamagata University, Yonezawa, Japan

**Abstract**

A simple method, *fishing* high-$T_c$ superconductor thin films out of liquid nitrogen bath by a permanent magnet (field > $H_{c1}$) due to the effect of high flux pinning, has been suggested to identify films having high critical current density ($J_c > 10^6$ A/cm$^2$ at 77 K) and thus a low microwave surface resistance ($R_s$). We have demonstrated that a Nd-Fe-B magnet, having a maximum field of ~ 0.5 T, could fish out Tl-1223 superconducting thin films on LSAT substrate with a thickness of ~ 5000 Å having $J_c > 1$ MA/cm$^2$ (at 77 K) whereas it could not fish out other films with $J_c < 0.1$ MA/cm$^2$ at 77 K. The fished out films exhibit $R_s$ values 237 – 245 µΩ at 77 K and 10 GHz, which is lower than that ($R_s = 317$ µΩ) of the best YBCO film at the same temperature and frequency. On the other hand, the non-fishable films show very high $R_s$ values. This method is a very simple tool to test for high $J_c$ and good microwave properties of superconducting films of large area which otherwise require a special and expensive tool.

The purpose of this article is to demonstrate a simple experiment to test superconducting films for high flux pinning and critical current density ($J_c$) and thus for a better microwave performance. The advent of superconductors with $T_c > 77$ K has made possible simple experiments with liquid nitrogen to test superconducting properties. The widely performed experiment of levitation of a magnet over a superconductor or vice versa represents superconducting state with or without high flux pinning.[1] On the other hand, the suspension of a superconductor below a magnet due to attractive forces represents solely the superconducting state with high flux pinning.[2] In other words, the simple test for high flux pinning is fishing or suspension rather than levitation of superconductors by a magnet. If there were no or low density of flux pinning present in the superconductor immersed in liquid nitrogen it would not be fished out or suspended by a magnet. We therefore adapted this experiment to examine flux pinning and $J_c$ of our Tl-1223 thin films, which are being developed for microwave device application.[3] High $J_c$ is an important factor for getting low microwave surface resistance.[4] In this report, we demonstrate that the fishing method can make clear distinction between high $J_c$ ( $> 1 \times 10^6$ A/cm$^2$) and low $J_c$ ($< 1 \times 10^5$ A/cm$^2$) of Tl-1223 thin films at 77 K and zero field and predict the microwave performance.

Fig. 1 shows a Tl-1223 film with a size of 20 mm × 20 mm Tl-1223 and a thickness of ~5000 Å on LSAT substrate fished out or suspended by a Ne-Fe-B magnet having a maximum field of ~ 0.5 T.[5] A total weight of two times that of the superconducting film with substrate could be fished out by adding another LSAT substrate on top of the superconducting film. It is intriguing that the attractive forces due to flux pinning in such a thin layer could lift a weight of 1000 times that of the thin superconducting film. This suggests that this film has high flux pinning and $J_c$. In fact, we have determined $J_c$ of a similar film with a size 10 mm × 10 mm after surface resistance measurements but cutting into 5 mm × 5 mm size by ac susceptibility method[6]. The $J_c$ value estimated at a dc field of 0.2 T was ~ $2 \times 10^5$ A/cm$^2$ at 77.3 K. We could not obtain $J_c$ value at zero field due to our experimental limitation of available ac field amplitude. However, a comparison of the behavior of $J_c$ versus dc field in other Tl-1223 film with smaller thickness and lower $J_c$, where the $J_c$ drops about one order with small applied magnetic fields (< 0.1 T),[6] it is expected that this film should have $J_c > 1 \times 10^6$ A/cm$^2$ at zero field. In our fishing test, some of the films could not be fished out. They were just repelled by the magnet. These films had $J_c$ about one order lower (<$1 \times 10^5$ A/cm$^2$) than that of the fished out films at 0.2 T. For better microwave properties, in addition to high $J_c$, the film should be homogeneous and free from defects.[7] Homogeneity and defects were examined by an optical microscope.[8]

Fig. 2 shows temperature dependence of microwave surface resistance of typical fishable and non-fishable Tl-1223 thin films. The surface resistance was measured by a HTS-sapphire-HTS resonator at 38 GHz.[9] It can be clearly seen that the values of $R_s$ of fishable Tl-1223 film labeled as A482 is much lower than those of the non-fishable film labeled as A436. The $R_s$ value of the fishable film at 77 K and 10 GHz is 245 μΩ according to $f^2$ law. This value is lower than that (317 μΩ) of the best YBCO film.[10] These results clearly suggest that the fishing method is a simple test for high $J_c$ and low $R_s$ superconducting thin films. Further, this method can form the basis for new simple instrument to measure $J_c$ of thin films, tapes and coated conductors.

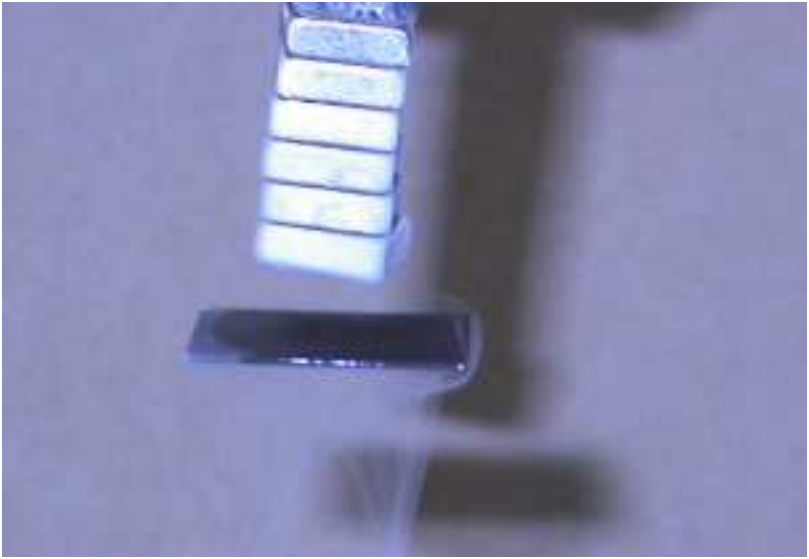

Fig. 1. Suspension of a Tl-1223 thin film with a size of 20 mm × 20 mm below a Nd-Fe-B magnet due to high flux pinning and $J_c$.

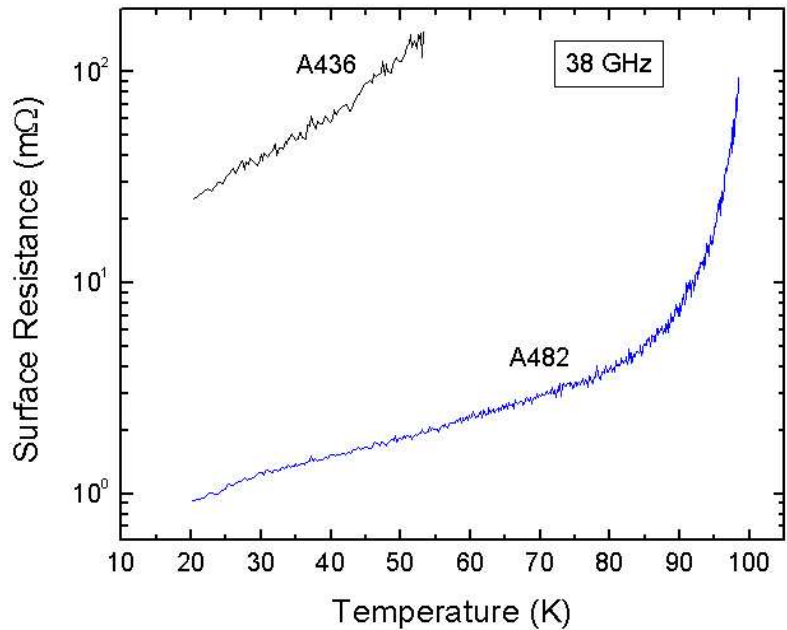

Fig. 2. Temperature dependence of microwave surface resistance of typical fishable (A482) and non-fishable (A436) Tl-1223 films at 38 GHz.

5